\documentclass[a4paper,11pt]{article}
\usepackage{cite}
\usepackage{amsmath}
\usepackage{amsthm}
\usepackage{amssymb}
\usepackage{amsfonts}
\usepackage[latin1]{inputenc}

\theoremstyle{theorem}
\newtheorem{theorem}{Theorem}
\newtheorem{proposition}[theorem]{Proposition}

\newtheorem{corollary}[theorem]{Corollary}

\theoremstyle{definition}
\newtheorem{definition}[theorem]{Definition}

\newtheorem{example}[theorem]{Example}

\theoremstyle{remark}
\newtheorem{remark}[theorem]{Remark}

\newcommand{\be}{\beta}

\newcommand{\ka}{\kappa}

\newcommand{\om}{\omega}

\newcommand{\vp}{\varphi}

\newcommand\Om\Omega
\newcommand\Te\Theta
\newcommand{\De}{\Delta}

\newcommand{\La}{\Lambda}

\newcommand{\bq}{\mathbf{q}}

\newcommand{\bA}{\mathbf{A}}
\newcommand{\bp}{\mathbf{p}}

\newcommand{\tH}{\widetilde{H}}

\newcommand{\tM}{\widetilde{M}}

\newcommand{\tga}{\widetilde{\gamma}}
\def\CC{\mathbb{C}}

\def\RR{\mathbb{R}}
\def\ZZ{\mathbb{Z}}

\renewcommand\SS{\mathbb S}
\newcommand{\cC}{{\mathcal C}}

\newcommand{\cE}{{\mathcal E}}
\newcommand{\cF}{{\mathcal F}}

\newcommand{\cM}{{\mathcal M}}

\newcommand{\cS}{{\mathcal S}}

\newcommand{\pd}{\partial}
\newcommand\minus\backslash
\newcommand{\id}{\rm id}

\newcommand\lan\langle
\newcommand\ran\rangle

\newcommand{\I}{{\mathrm i}}
\newcommand{\e}{{\mathrm e}}
\newcommand{\dd}{{\mathrm d}}
\newcommand{\1}{{\mathrm{I}}}
\newcommand{\2}{{\mathrm{II}}}
\newcommand\J{j=\text{I,\,II}}
\newcommand\K{_{\rm K}} \newcommand\HO{_{\rm H}}

\title{\Large\bf Hamiltonian systems admitting a Runge--Lenz vector and an optimal extension of Bertrand's theorem to curved manifolds}
\author{\normalsize Ángel Ballesteros$^a$\thanks{angelb@ubu.es},\;
Alberto Enciso$^b$\thanks{aenciso@fis.ucm.es},\; Francisco J.
Herranz$^a$\thanks{fjherranz@ubu.es},\; Orlando
Ragnisco$^c$\thanks{ragnisco@fis.uniroma3.it}}
\date{\small $^a$ Depto.\ de Física, Universidad de Burgos,
09001 Burgos, Spain\vspace{1ex}\\
$^b$ Depto.~de F{í}sica Teórica II, 
Universidad Complutense, 
28040 Madrid, Spain\vspace{1ex}\\
$^c$ Dip.\ di Fisica, Università di Roma 3, and Istituto
Nazionale di Fisica Nucleare,
\\ 00146 Rome, Italy}

\begin{document}
\maketitle

\begin{abstract}
Bertrand's theorem asserts that any spherically symmetric natural
Hamiltonian system in Euclidean 3-space which possesses stable
circular orbits and whose bounded trajectories are all periodic is
either a harmonic oscillator or a Kepler system. In this paper we
extend this classical result to curved spaces by proving that any Hamiltonian on a spherically symmetric Riemannian
3-manifold which satisfies the same conditions as in Bertrand's
theorem is superintegrable and given by an intrinsic
oscillator or Kepler system. As a byproduct we obtain a wide
panoply of new superintegrable Hamiltonian systems. The demonstration
relies on Perlick's classification of Bertrand spacetimes and on the
construction of a suitable, globally defined generalization of the
Runge--Lenz vector.\\[3mm]
\end{abstract}

\section{Introduction and preliminary definitions}
\label{S:intro}

The Kepler problem and the harmonic oscillator are probably the most thoroughly studied systems in classical mechanics. The reasons for this are twofold. First, these potentials play a preponderant role in Physics due their connection with planetary motion and oscillations
around a nondegenerate equilibrium. Second, these potentials are of particular mathematical interest due to the existence of additional (or ``hidden'') symmetries yielding additional constants of motion. In fact, both the Kepler and the harmonic oscillator Hamiltonians are (maximally) {\em superintegrable} in the sense that they have the maximum number (four) of functionally independent first integrals other than the Hamiltonian.\footnote{As usual, by functional independence of the integrals $I_1,\dots, I_k$ we mean that the $(k+1)$-form $\dd H\wedge\dd I_1\wedge\dots\wedge\dd I_k$ is nonzero in an open and dense subset of phase space, $H$ being the Hamiltonian function.}

Bertrand's theorem~\cite{Be73} is a landmark result which characterizes the Kepler and harmonic oscillator Hamiltonians in terms of their qualitative dynamics. A precise statement of this theorem is given below. We recall~\cite{Fe04} that the first condition, which is occasionally forgotten, is necessary in order to exclude potentials of the form $V(\bq)=-K\|\bq\|^{-s}$, with $K>0$ and $s=2,3,\dots$

\begin{theorem}[Bertrand]\label{T:Bertrand}
Let $H=\frac12\|\bp\|^2+V(\bq)$ be a natural, spherically symmetric Hamiltonian system in a domain of $\RR^3$. Let us suppose that:
\begin{enumerate}
\item There exist stable circular orbits.

\item All the bounded trajectories are closed.
\end{enumerate}
Then the potential is either a Kepler $(V(\bq)=A/\|\bq\|+B)$ or a harmonic oscillator potential $(V(\bq)=A\|\bq\|^2+B)$. In particular, $H$ is superintegrable.
\end{theorem}

Analogues of the Kepler and harmonic oscillator systems in curved spaces have been of interest since the discovery of non-Euclidean geometry. In fact~\cite{Sh05}, the ``intrinsic'' Kepler and harmonic oscillator problems on spaces of constant curvature were studied by Lipschitz and Killing already in the 19th century, and later rediscovered by Schrödinger~\cite{Sc40} and Higgs~\cite{Hi79}. In both cases it was established that these systems are superintegrable and satisfy Properties (i) and (ii) above.

A considerably more ambitious development was Perlick's introduction and classification of Bertrand spacetimes~\cite{Pe92}, which was based on the following observation. Let $(M,g)$ be a Riemannian 3-manifold and consider the space $\cM=M×\RR$ endowed with the warped Lorentzian metric $\eta=g-\frac1{V}\dd t^2$, with $V$ a smooth positive function on $M$. Then the {\em trajectories} in $(\cM,\eta)$, that is, the projections of inextendible timelike geodesics to a constant time leaf $M×\{t_0\}$, correspond to integral curves of the Hamiltonian $H=\frac12\|p\|_{g}^2+V(q)$ in (the cotangent bundle of) $M$. Thus Perlick introduced the following

\begin{definition}\label{D:Bertrand}
A Lorentzian $4$-manifold $(M×\RR,\eta)$ is a {\em Bertrand spacetime} if:
\begin{enumerate}
\item It is spherically symmetric and static in the sense that $\eta=g-\frac1V\dd t^2$ and $M$ is diffeomorphic to $(r_1,r_2)×\SS^2$, where the smooth function $V$ depends only on $r$ and the Riemannian metric $g$ on $M$ takes the form
\begin{equation}\label{metricB}
g=h(r)^2\,\dd r^2+r^2\big(\dd\theta^2+\sin^2\theta\,\dd\varphi^2\big)
\end{equation}
in the adapted coordinate system $(r,\theta,\vp)$. Here $r_1,r_2\in\RR^+\cup\{+\infty\}$.

\item There is a circular ($r=\text{const.}$) trajectory passing through each point of $M$.

\item The above circular trajectories are stable, that is, any initial condition sufficiently close to that of a circular trajectory gives a periodic trajectory.
\end{enumerate}
\end{definition}

Perlick's main result was the classification of all Bertrand spacetimes, recovering the classical Bertrand theorem as a subcase. However, two main related questions remained to be settled. On the one hand, the {\em potentials} $V$ in Perlick's classification lacked any physical interpretation, and this was in strong contrast with the Euclidean case. This drawback was circumvented in Ref.~\cite{BEHR08}, where we showed that the two families of Perlick's potentials correspond to either the ``intrinsic'' Kepler or harmonic oscillator potentials in the underlying 3-manifold $(M,g)$. On the other hand, the issue of whether the corresponding Hamiltonian systems were superintegrable in some reasonable sense was left wide open. In fact, Perlick's only remark in this direction was that, by virtue of a theorem of Hauser and Malhiot~\cite{HM74}, only two concrete models among the family of Bertrand spacetimes admitted a quadratic additional integral coming from a second rank Killing tensor.

A careful analysis of the literature reveals that many particular cases of Bertrand metrics have been thoroughly analyzed and shown to be superintegrable~\cite{GR88,IK94,IK95,Gi07}, and that in many cases they have been shown to admit a generalization of the classical Runge--Lenz vector as an additional first integral. The physical and mathematical interest of these models (and thus of Bertrand spacetimes) is fostered by their connections with the theory of magnetic monopoles, with differential and algebraic geometry, and with low-dimensional manifold theory~\cite{Br89,SW94,KM97,CK01,KMOZ07,KN07,NR07,Na08,NY08}. The relation between Bertrand spaces and monopole motion is not totally incidental. Indeed, an ample subclass of Bertrand spacetimes admitting some kind of generalized Runge--Lenz vectors (the so-called multifold Kepler systems) were introduced by Iwai and Katayama~\cite{IK94,IK95} as generalizations of the Taub--NUT metric, whose geodesics asymptotically describe the relative motion of two monopoles (see, for instance,~\cite{Mac70,Ma82,AH85, CFH90, BCJM, JL}). Interestingly, superintegrable Hamiltonian systems on curved spaces have recently attracted considerable attention also within the integrable systems community, especially in low dimensions (cf.~\cite{KKMW02,BHSS03,KKMW03,KKM06,BH07,BEHR08a} and references therein).

The main result of this article is that all Bertrand spacetimes are indeed superintegrable, their superintegrability being linked to the existence of a generalized Runge--Lenz vector. This enables us to present an optimal version of Bertrand's theorem (Theorem~\ref{T:main}) on spherically symmetric manifolds which includes the classification of the natural Hamiltonians whose bounded orbits are all periodic~\cite{Pe92}, the physical interpretation of the corresponding potentials as Kepler or harmonic oscillator potentials, in each case, and the proof of the superintegrability of these models. This settles in a quite satisfactory way a problem with a large body of previous partial results scattered in the literature.

It is standard that the superintegrability of the Kepler system stems from the existence a conserved Runge--Lenz vector, whose geometric significance is described from a modern perspective in~\cite{MZ}. On the other hand, the superintegrability of the harmonic oscillator is usually established either using explicit (scalar) first integrals or the conserved rank 2 tensor $\mathbf{C}=2\om^2\bq\otimes\bq+\bp\otimes\bp$, which is sometimes preferable for algebraic reasons~\cite{Fr65}. That the latter approach is closely related to a (multivalued) analogue of the Runge--Lenz vector was firmly established in~\cite{HM90}. Motivated by this connection, we have based our approach to the integrability of the Bertrand systems on the construction of a generalized Runge--Lenz vector, globally defined on a finite cover of $M$. This construction relies on a detailed analysis of the integral curves of the appropriate Hamiltonians. The literature on generalizations of the Runge--Lenz vector for central potentials on Euclidean space is vast (see the survey~\cite{LF03} and references therein), but unfortunately several interesting papers are severely flawed by the lack of distinction between local, semi-global and global existence.

The article is organized as follows. In Section~\ref{S:potentials} we recall the two families of Bertrand spacetimes entering Perlick's classification, which are labeled by two coprime positive integers $n$ and $m$. We also include the characterization of Perlick's potentials as the intrinsic Kepler or harmonic oscillator potentials of the corresponding Riemannian 3-manifolds $(M,g)$ and briefly discuss several physically relevant examples. In Section~\ref{S:superint1} we consider the associated natural Hamiltonian systems on $(M,g)$ and compute their integral curves in closed form (Proposition~\ref{P:orbits}). Using this result we easily derive that the latter Hamiltonians are geometrically superintegrable (cf.\ Definition~\ref{D:geom.superint} and Proposition~\ref{P:superint1}) in the region of phase space foliated by bounded orbits, as happens with the harmonic oscillator and Kepler potentials in $\RR^3$. Our central result is a stronger superintegrability theorem (Theorem~\ref{T:Runge}) that we present in Section~\ref{S:Runge}, where we construct a generalized Runge--Lenz vector globally defined on an $n$-fold cover of $M$. As a corollary of this construction we also obtain a global rank $n$ tensor field in $M$ invariant under the flow and a wide panoply of new superintegrable Hamiltonian systems. Lastly, in Section~\ref{S:examples} we combine the results established in the previous sections to obtain an optimal extension of Bertrand's theorem to curved spaces (Theorem~\ref{T:main}).

\section{Harmonic oscillators and Kepler potentials in Bertrand spacetimes}
\label{S:potentials}

In this section we shall define the ``intrinsic'' Kepler and harmonic oscillator potentials in a spherically symmetric 3-manifold and show how Bertrand spacetimes are related to the Kepler and harmonic oscillator potentials of any of its constant time leaves. Most of the material included here is essentially taken from Ref.~\cite{BEHR08}; for the sake of completeness, let us mention that further information on geometric properties of Green functions can be consulted e.g.\ in~\cite{LT87,LT95,EP07,EP08,EP09}.

We start by letting $(M,g)$ be a Riemannian 3-manifold as in Definition~\ref{D:Bertrand}. In particular, the metric $g$ takes the form
\begin{equation}\label{metric3gen}
\dd s^2=h(r)^2\,\dd r^2+r^2\big(\dd\theta^2+\sin^2\theta\,\dd\varphi^2\big)\,.
\end{equation}
It is standard that if $u(r)$ is function which depends only on the radial coordinate, then its Laplacian is also radial and reads as
\[
\De_gu(r)=\frac{1}{r^2h(r)}\frac{\dd}{\dd r}\bigg(\frac{r^2}{h(r)}\frac{\dd u}{\dd r}\bigg)\,.
\]
As the Kepler potential in Euclidean three-dimensional space is simply the radial Green function of the Laplacian and the harmonic oscillator is its inverse square, it is natural to make the following

\begin{definition}\label{D:KHO}
The (intrinsic) {\em Kepler} and the {\em harmonic oscillator potentials} in $(M,g)$ are respectively given by the radial functions
\begin{equation}\label{KHO}
V\K(r)=A_1\bigg(\int^r_ar'^{-2}h(r')\,\dd r'+B_1\bigg)\,,\qquad V\HO(r)=A_2\bigg(\int^r_ar'^{-2}h(r')\,\dd r'+B_2\bigg)^{-2}\,,
\end{equation}
where $a,A_j,B_j$ are constants.
\end{definition}

\begin{example}\label{E:CCS}
Let $(M,g)$ be the simple connected, three-dimensional space form of sectional curvature $\ka$. In this case the metric has the form~\eqref{metric3gen} with
\[
h(r)^2=\frac1{1-\ka r^2}\,.
\]
The corresponding Kepler and harmonic oscillator potentials are therefore
\begin{equation}\label{KHOCCS}
V\K=\sqrt{r^{-2}-\ka}\,,\qquad V\HO=\frac1{r^{-2}-\ka}
\end{equation}
up to additive and multiplicative constants. In terms of the distance function $\rho_\ka$ to the point $r=0$ this can be rewritten as
\[
V\K=\sqrt\ka\cot\big(\sqrt\ka\,\rho_\ka\big)\,,\qquad V\HO=\frac{\tan^2(\sqrt\ka\,\rho_\ka)}\ka\,,
\]
thus reproducing the known prescriptions for the sphere and the hyperbolic space~\cite{Sh05,BH07}. The Euclidean case is recovered by letting $\ka\to0$.
\end{example}

Now let us consider the spherically symmetric spaces $(M,g_j)$ ($\J$) defined by the metrics
\begin{subequations}\label{3metric}
\begin{align}
{\mbox {Type  I} }: \quad& \dd
s^2=  \frac{m^2\dd r^2 }{n^2\left( 1+Kr^2\right) }
+r^2(\dd\theta^2+\sin^2\theta\,\dd\varphi^2)\\
{\mbox {Type  II}}:\quad & \dd
s^2=\frac{2m^2\left( 1-Dr^2±\sqrt{(1-Dr^2)^2-Kr^4}
\right)}{n^2\left((1-Dr^2)^2-Kr^4
\right)}\,\dd r^2 +r^2(\dd\theta^2+\sin^2\theta\,\dd\varphi^2)
\end{align}
\end{subequations}
where $D$ and $K$ are real constants and $m$ and $n$ are coprime positive integers. The  maximal interval $(r_1,r_2)$ can be easily found from these expressions. These Riemannian 3-manifolds, which first appeared in~\cite{Pe92} (where the quotient $n/m$ was called $\be$), will be henceforth called {\em Bertrand spaces}. A short computation shows that, up to a multiplicative constant, the Kepler potential of a Bertrand space of type I is
\begin{subequations}\label{Vs}
\begin{equation}\label{V1}
V_\1=\sqrt{ r^{-2} +K} +G\,,
\end{equation}
whereas the harmonic oscillator potential of one of type II can be written in the convenient form
\begin{equation}\label{V2}
V_\2=G\mp r^2\Big(1-Dr^2±\sqrt{(1-Dr^2)^2-Kr^4}\Big)^{-1}\,.
\end{equation}
\end{subequations}
Here $G$ is an arbitrary constant.

By comparing with Ref.~\cite{Pe92}, the above digression immediately yields the following

\begin{proposition}\label{P:potentials}
$(\cM,\eta)$ is a Bertrand spacetime if and only if it is isometric to the warped product $(M×\RR,g_j-\frac{\dd t^2}{V_j})$, with $(M,g_j)$ a Bertrand space of type $j$ ($\J$, cf.~\eqref{3metric}) and $V_j$ given by~\eqref{Vs}.
\end{proposition}

In particular, this shows that Perlick's obtention of two different kinds of Bertrand spacetimes has a natural interpretation~\cite{BEHR08}: they are associated to either Kepler (type I) or harmonic oscillator (type II) potentials. The multiplicative constant of the potentials is inessential and can be eliminated by rescaling the time variable.

\begin{example}\label{E: ex}
We conclude this section with a brief discussion of a few examples of physically relevant spaces that are Bertrand. This intends both to serve as motivation and to help the reader gain some insight on Bertrand spaces. A more detailed discussion can be found in~\cite{BEHR08}.

\begin{enumerate}
\item {\em Spaces of constant curvature.} The metric of the simply connected Riemannian 3-manifold of constant sectional curvature $\ka$ is usually written as
\[
\dd s^2=\frac{ \dd r^2 }{ 1-\ka r^2}
+r^2\big(\dd\theta^2+\sin^2\theta\,\dd\vp^2\big)\,.
\]
We have already seen that the Kepler and harmonic oscillator potentials in these spaces are given by~\eqref{KHOCCS}, and it is well known that all the bounded integral curves of both systems are periodic. This result is immediately recovered by noticing that the Kepler system is recovered from the type I Bertrand spacetimes when $n=m=1$ and $K=-\ka$, whereas the harmonic oscillator is obtained as the type II Bertrand spacetime with $n/m=2$, $K=0$ and $D=\ka$.

\item {\em Darboux space of type III.} Consider the metric
\[
\dd s^2= \frac{k^2+2 r^2+k \sqrt{k^2+4 r^2}}{2(k^2+4 r^2)}\,\dd r^2+r^2\big(\dd\theta^2+\sin^2\theta\,\dd\vp^2\big)\,,
\]
whose intrinsic harmonic oscillator potential is given by
\[
V_\2=\frac{2 k^2 r^2}{k^2+2 r^2+k \sqrt{k^2+4 r^2}}
\]
up to multiplicative and additive constants. This defines a Bertrand spacetime of type II with parameters $n/m=2$, $K=4/k^4$ and $D=-2/k^2$.

Let us introduce coordinates $\mathbf Q=(Q^1,Q^2,Q^3)$ as
\[
\mathbf Q=\bigg(\frac{(k^2+4r^2)^{1/2}-k}2\bigg)^{1/2}\big(\cos\theta\,\cos\vp,\cos\theta\,\sin\vp,\sin\theta\big)\,.
\]
In terms of these coordinates, the above metric and potential read as
\[
\dd s^2=\big(k+\|\mathbf Q\|^2\big)\|\dd\mathbf Q\|^2\,,\qquad V_\2=\frac{k^2\|\mathbf Q\|^2}{k+\|\mathbf Q\|^2}\,.
\]
Thus we recover the three-dimensional Darboux system of type III~\cite{KKMW03}. The Darboux system of type III is the only quadratically superintegrable natural Hamiltonian system in a surface of nonconstant curvature which is known to admit quadratically superintegrable $N$-dimensional generalizations~\cite{BEHR08a}.

\item {\em Multifold Kepler systems.} The family of multifold Kepler systems was introduced by Iwai and Katayama~\cite{IK94,IK95} as Hamiltonian reductions of the geodesic flow in a generalized Taub--NUT metric. These systems are given by the metrics and potentials
\begin{align*}
\dd s^2&=\|\mathbf Q\|^{\frac  nm -2} \big(a + b\,\|\mathbf Q\|^{\frac nm}\big)  \|\dd \mathbf Q\|^2\,,\\
V_\2&= \frac { \|\mathbf Q\|^{2- \frac nm} }{ a + b \|\mathbf Q\|^{\frac nm}}\left(
\mu^2 \|\mathbf Q\|^{-2}+ \mu^2 c\, \|\mathbf Q\|^{\frac nm -2}+ \mu^2 d\,\|\mathbf Q\|^{\frac {2m}n -2} \right)\,,
\end{align*}
with $\mathbf Q=(Q^1,Q^2,Q^3)$, $a,b,c,d,\mu$ constants and $n,m$ coprime positive integers. The substitution
\[
\mathbf Q=\bigg(\frac{(a^2+4br^2)^{\frac12}-a}{2b}\bigg)^{\frac mn}\big(\cos\theta\,\cos\vp,\cos\theta\,\sin\vp,\sin\theta\big)
\]
shows that the multifold Kepler models are equivalent to the type II Bertrand systems with parameters $K = 4a^{-4}b^2$ and $D= -{\frac{2b}{a^2}}$. It should be noticed that the Darboux space of type III is a particular case of the multifold Kepler systems.

\end{enumerate}  
\end{example}

\section{The orbit equation and geometric superintegrability}
\label{S:superint1}

Hereafter we shall analyze the properties of the Hamiltonian systems in $(M,g_j)$ given by
\begin{equation}\label{H}
H_j:=\frac12\|p\|^2_{g_j}+V_j(q)\,,\qquad \J\,,
\end{equation}
where the metric $g_j$ and the potential $V_j$ are respectively defined by~\eqref{3metric} and~\eqref{Vs}. As previously discussed, the orbits of these systems correspond to trajectories of the associated Bertrand spacetimes. It should be noticed that in the adapted coordinate system, these Hamiltonians read as
\begin{subequations}\label{Hcoord}
\begin{align}
H_\1&=\frac12\Bigg[\Big(\frac{n}{m}\Big)^2\left( 1+Kr^2\right)p_r^2
+\frac{p_\theta^2}{r^2}+\frac{p_\varphi^2}{r^2\sin^2\theta}\Bigg]+\sqrt{ r^{-2} +K} +G\,,\label{H1}\\
H_\2&=\frac12\Bigg[\frac{n^2\left((1-Dr^2)^2-Kr^4
\right)p_r^2}{2m^2\left( 1-Dr^2±\sqrt{(1-Dr^2)^2-Kr^4}
\right)} +\frac{p_\theta^2}{r^2}+\frac{p_\varphi^2}{r^2\sin^2\theta}\Bigg]\notag\\
&\qquad\qquad\qquad\qquad\qquad\mp r^2\Big(1-Dr^2±\sqrt{(1-Dr^2)^2-Kr^4}\Big)^{-1}+G\,,\label{H2}
\end{align}
\end{subequations}
where $p_r$ is the momentum conjugate to $r$ and $p_\theta$ and $p_\vp$ are defined analogously.

In this section we shall derive the simplest superintegrability property of the Hamiltonian systems~\eqref{H} (cf.\ Proposition~\ref{P:superint1}), which nonetheless seems to have escaped unnoticed so far. The proof of this result relies on the fact that, by definition, the orbits of~\eqref{H} define an invariant foliation by (topological) circles in an open subset $\Om\subset T^*M$ of the phase space of the system. E.g., in the classical Kepler problem
\[
\Om=\big\{(\bq,\bp)\in\RR^3×\RR^3:H(\bq,\bp)<0,\; \bq×\bp\neq0\big\}
\]
is the set of points with negative energy and nonzero angular momentum,
whereas for the harmonic oscillator one can take $\Om=(\RR^3×\RR^3)\minus\{(0,0)\}$, i.e., the whole phase space minus the equilibrium. In Proposition~\ref{P:orbits} below we compute the expression of the orbits in closed form, revealing that the above foliation is actually a locally trivial fibration. This allows us to resort to the geometric theory of superintegrable Hamiltonian systems~\cite{DD87}, yielding the first superintegrability result for~\eqref{H}.

Before discussing the precise statement of Proposition~\ref{P:superint1}, let us compute the orbits of the Hamiltonian~\eqref{H}. In fact, the closed expression that we shall derive is not only used in the proof of Proposition~\ref{P:superint1}, but it is also a key element of Theorem~\ref{T:Runge}, where a stronger superintegrability result is presented. It is convenient to introduce the rectangular coordinates $\bq=(q^1,q^2,q^3)$ associated to the spherical coordinates $(r,\theta,\vp)$ as
\begin{equation}\label{bq}
\bq=\big(r\cos\theta\cos\vp,r\cos\theta\sin\vp,r\sin\theta\big)\,.
\end{equation}
The conjugate momenta will be denoted by $\bp=(p_1,p_2,p_3)$. Clearly the coordinates $(\bq,\bp)$ are globally defined in $T^*M$. We shall use the notation $\cdot$, $×$ and $\|\cdot\|$ for the Euclidean inner product, cross product and norm in $\RR^3$ and call $E=H_j(\bp,\bq)$ and $J^2=\|\bq×\bp\|^2$ the energy and angular momentum of an integral curve $(\bq(t),\bp(t))$ of~\eqref{H}. Obviously $E$ and $J^2$ are constants of motion.

\begin{proposition}\label{P:orbits}
Let $\gamma$ be an inextendible orbit of the Hamiltonian system~\eqref{H} which is contained in the invariant plane $\{\theta=\frac\pi2\}$. Then $\gamma$ is given by
\begin{subequations}\label{orbits}
\begin{equation}\label{orbit1}
\cos\bigg(\frac{n\vp}m-\vp_0\bigg)=\frac{1+J^2\sqrt{r^{-2}+K}}{\sqrt{1+2J^2(E-G)+KJ^4}}
\end{equation}
if $j=\1$ and by
\begin{equation}\label{orbit2}
\cos\bigg(\frac{n\vp}m-\vp_0\bigg)=\frac{J^2r^{-2}\big(1-D r^2±\sqrt{(1-D r^2)^2-Kr^4}\big)+DJ^2+2G-2E}{\sqrt{(2E-2G-DJ^2)^2±4J^2-KJ^4}}
\end{equation}
\end{subequations}
if $j=\2$. Here $\vp_0$ is a real constant.
\end{proposition}
\begin{proof}
We begin with the case $j=\1$. The crucial observation is that the orbit equation
\[
\frac{m^2J^2}{n^2r^4(1+Kr^2)}\bigg(\frac{\dd r}{\dd\vp}\bigg)^2=2E-2V_\1-\frac{J^2}{r^2}
\]
simplifies dramatically with the change of variables
\[
u=\sqrt{r^{-2}+K}\,,
\]
in terms of which the potential and the inverse square term read as
\[
V_\1=u+G\,,\qquad r^{-2}=u^2-K\,.
\]
The orbit equation is then given by
\[
\bigg(\frac{mJ}{n}\frac{\dd u}{\dd\vp}\bigg)^2=2E-2G+KJ^2-2u-J^2u^2\,,
\]
which can be readily integrated to yield
\[
\cos\bigg(\frac{n\vp}m-\vp_0\bigg)=\frac{1+J^2u}{\sqrt{1+2J^2(E-G)+KJ^4}}
\]
for some constant $\vp_0$.

When $j=\2$ the treatment is analogous.  Now the orbit equation reads
\[
\frac{1-Dr^2±\sqrt{(1-Dr^2)^2-Kr^4}}{r^4\big[(1-Dr^2)^2-Kr^4\big]}
\bigg(\frac{mJ}{n}\frac{\dd r}{\dd\vp}\bigg)^2=E-V_\2-\frac{J^2}{2r^2}\,,
\]
and it is convenient to introduce the variable
\[
v=r^{-2}\Big(1-D r^2±\sqrt{(1-D r^2)^2-Kr^4}\Big)\,.
\]
In terms of this new coordinate the potential is simply $V_\2=G\mp\frac1v$, whereas the inverse square term is given by
\[
r^{-2}=\frac{v^2+2Dv+K}{2v}\,.
\]
Hence a straightforward computation shows that the orbit equation is
\[
\bigg(\frac{mJ}{n}\frac{\dd v}{\dd\vp}\bigg)^2=4(E-G)v-J^2\big(v^2+2Dv+K\big)±4\,,
\]
thereby obtaining
\[
\cos\bigg(\frac{n\vp}m-\vp_0\bigg)=\frac{J^2(v+D)+2G-2E}{\sqrt{(2E-2G-DJ^2)^2±4J^2-KJ^4}}\,.
\]
Here $\vp_0$ is a real constant.
\end{proof}
\begin{remark}\label{R:J=0}
Eqs.~\eqref{orbits} are well defined also when $J=0$. Moreover, it is not difficult to check that $r$ can be readily expressed as a function of $\vp$ by performing some manipulations in the right-hand side of~\eqref{orbits}.
\end{remark}

We shall now specify what is understood by geometric superintegrability.  Let $F_0$ be a smooth Hamiltonian defined on a $2d$ dimensional symplectic manifold $N$ admitting $s\geq d-1$ functionally independent first integrals $F_1,\dots, F_{s}$ other than the Hamiltonian. Let us suppose that $F=(F_0,F_1,\dots, F_s)$ is a submersion onto its image with compact and connected fibers, which by Ehresmann's theorem~(cf.\ e.g.~\cite{Me02}) implies that its level sets define a locally trivial fibration $\cF$ of $N$. If $s\geq d$, not all the latter first integrals can Poisson-commute: the usual condition to impose is that there exists a matrix-valued function $P:F(N)\to{\rm Mat}(s+1)$ of rank $s-d+1$ such that
\begin{equation}\label{rank}
\{F_i,F_j\}=P_{ij}\circ F\,,\qquad 0\leq i,j\leq s\,.
\end{equation}
In particular, when $s=d-1$ this yields the usual definition of Liouville integrability. Well known generalizations of the Liouville--Arnold theorem~\cite{Ne72,MF78} show that every fiber of $F$ is an invariant $(2d-s-1)$-torus, and that the motion on each of these tori is conjugate to a linear flow. Moreover, the fibration $\cF$ has symplectic local trivializations.

Geometrically, the existence of the function $P$ means that $\cF$ has a polar foliation~\cite{DD87}, i.e, a foliation $\cF^\perp$ whose tangent spaces are symplectically orthogonal to those of $\cF$. Similarly, the rank condition in Eq.~\eqref{rank} is tantamount to demand that the invariant $(2d-s-1)$-tori of the foliation be isotropic. Thus the crucial element in the geometric characterization of superintegrability is the bifoliation $(\cF,\cF^\perp)$, which is a type of dual pair as defined in~\cite{We83}. One is thus led to introduce the following definition (cf.~\cite{DD87} and the survey~\cite{Fa05}, where slightly different wording is used):

\begin{definition}\label{D:geom.superint}
A Hamiltonian system on a symplectic $2d$-dimensional manifold is {\em geometrically superintegrable with $s\geq d-1$ semiglobal integrals} if the Hamiltonian vector field is tangent to a locally trivial fibration by isotropic $(2d-s-1)$-tori which admits a polar foliation. If $s$ takes the maximum value $2d-2$ we shall simply say that the system is {\em geometricaly superintegrable}.
\end{definition}

Of course, generally not all the phase space of a (super)integrable system is fibered by invariant isotropic tori: there can be, e.g., singular points and unbounded orbits. But it is customary and of interest to restrict one's attention to the region where such fibration is well defined. In the case when $s=d-1$ (Liouville integrability), the invariant tori are Langrangian and therefore $\cF^\perp=\cF$, explaining why the bifibration $(\cF,\cF^\perp)$ is less well known than the fibrations by Lagrangian tori. (However, an advantage of the bifibration is that, under mild technical assumptions, it is uniquely determined (and finer), whereas for integrable systems with additional integrals there is some arbitrariness in the choice of invariant Lagrangian tori.) It should be noticed that the above structure yields ``semiglobal'' (i.e., defined in a tubular neighborhood of each torus) first integrals associated to the existence of generalized action-angle coordinates; a detailed account can be found in~\cite{DD87,Bo96,Fa05}. The content of the following proposition is that the Bertrand systems~\eqref{H} are geometrically superintegrable in the region foliated by periodic orbits.

\begin{proposition}\label{P:superint1}
Let $\Om$ be the region of $T^*M$ where all the orbits of $H_j$ are periodic. Then $H_j|_\Om$ is geometrically superintegrable.
\end{proposition}
\begin{proof}
It easily follows from Proposition~\ref{P:orbits} that the orbits of $H_j$ define a locally trivial fibration by (topological) circles in $\Om$. The fibers are certainly isotropic, as they are one-dimensional, and the flow of $H_j$ on each fiber is conjugate to the linear one because $H_j$ does not possess any critical points in $\Om$. Moreover, it stems from Proposition~\ref{P:orbits} that the function $\Om\to\RR^+$ mapping each point in $\Om$ to the length of the (periodic) orbit passing through it is smooth, which in turn readily implies that the period function is also smooth and nonvanishing in this region. Hence a theorem of Fassò~\cite{Fa98} implies that $H_j$ is geometrically superintegrable, proving the proposition.
\end{proof}

\section{The generalized Runge--Lenz vector}
\label{S:Runge}

In this section we shall prove a stronger superintegrability result for the Bertrand Hamiltonians~\eqref{H}. More precisely, we shall provide a semi-explicit construction of an additional vector first integral which we shall call the generalized Runge--Lenz vector. This vector field is defined on an $n$-fold cover $\tM$ of the original space $M$, and it is invariant under the flow generated by the lift of the Bertrand Hamiltonian to the covering space $\tM$. In $M$, this vector field induces a global tensor field of rank $n$ which is preserved under the flow of $H$. As before, $n$ is the positive integer which appears in Eq.~\eqref{H}.

As regards the superintegrability properties of the Hamiltonian systems~\eqref{H}, the spherical symmetry of these systems readily yields three first integrals other than the Hamiltonian, which can be identified with the components of the angular momentum. The idea of looking for generalizations of the Runge--Lenz vector in order to find an additional integral of motion is not new: an updated and rather complete review of the related literature can be found in~\cite{LF03}. Here we shall use our information about the integral curves of~\eqref{H} and some ideas already present in the work of Fradkin~\cite{Fr67} and Holas and March~\cite{HM90}.

Let us start by recalling Fradkin's construction~\cite{Fr67} of a {\em local} vector first integral for the Hamiltonian system
\[
H_0=\frac12\|\bp\|^2+U(\|\bq\|)\,,
\]
where $U(\|\bq\|)$ is an arbitrary central potential and $(\bq,\bp)\in\RR^3×\RR^3$. The starting point is the following trivial remark. Consider an integral curve $\bq(t)$ of $H_0$ contained in the plane $\{\theta=\frac\pi2\}\subset\RR^3$, where $(r,\theta,\vp)$ are the usual spherical coordinates. We can assume without loss of generality that we have taken the initial condition $\vp(0)=0$ and use the notation $r=\|\bq\|$, $J=p_\varphi=r^2\dot\vp$. A simple computation shows that the derivative along this integral curve of the unit vector field
\begin{equation}\label{Fradkin}
{\mathbf a}=\frac{\cos\vp}r\,\bq+\frac{\sin\vp}{rJ}\,\bq×(\bq×\bp)
\end{equation}
is identically zero, as in fact ${\mathbf a}(t)$ is the constant vector $(1,0,0)$. Fradkin's observation was that if $\cos\vp$ and $J^{-1}\sin\vp$ can be expressed in terms of $\bq$ and $\bp$ in a domain $\Om\subset\RR^3\minus\{0\}$, then the resulting vector field is a first integral of $H_0$ in $\Om$. When $H_0$ is the Kepler Hamiltonian, the generalized Runge--Lenz vector field is well defined globally and essentially coincides with the classical Runge--Lenz vector divided by its norm. When $H_0$ is the harmonic oscillator, the generalized Runge--Lenz vector is multivalued (this can be neatly understood by considering the turning points of the orbits), but can be used to recover the conserved tensor field $\mathbf{C}=2\om^2\bq\otimes\bq+\bp\otimes\bp$ associated to the $\mathrm{SU}(3)$ symmetry~\cite{HM90}.

\begin{definition}\label{D:runge}
Let $H$ be a Hamiltonian system defined on (the cotangent bundle of) a 3-manifold $N$. We say that $H$ admits a {\em generalized Runge--Lenz vector} if there exists a nontrivial horizontal vector field $A$ in $T^*N$ which is constant along the flow of $H$.
\end{definition}

Obviously the conserved vector $A$ is nontrivial if it is not constant and cannot be written in terms of the energy and the angular momentum integrals, and we recall that a horizontal vector (resp.\ tensor) field in $T^*N$ can be simply understood as a vector (resp.\ tensor) field in $N$ which depends on both the positions and the momenta. The main problem with Fradkin's approach is that, of course, it is not at all obvious how to obtain sufficient conditions ensuring that these local integrals are well defined globally, while local superintegrability is trivial in a neighborhood of any regular point of the Hamiltonian flow. However, we shall see below that Fradkin's approach works well for the kind of Hamiltonian systems that we are considering in this paper, and that one can construct a globally defined generalized Runge--Lenz vector (cf.\ Eq.~\eqref{Ak} below) which is roughly analogous to~\eqref{Fradkin}.

\begin{theorem}\label{T:Runge}
Consider a Hamiltonian of the form~\eqref{H}, with $m,n$ coprime positive integers. Then there exists an $n$-fold cover $\tM$ of $M$ such that the lift of this Hamiltonian to $\tM$ admits a generalized Runge--Lenz vector.
\end{theorem}
\begin{proof}
We shall call $H_j$, $\J$, the Hamiltonian~\eqref{H}. Let $\gamma$ be an inextendible orbit of $H_j$, which can be assumed to lie in the invariant plane $\{\theta=\frac\pi2\}$. By Proposition~\ref{P:orbits}, and taking $\vp_0=0$ in Eq.~\eqref{orbits} without loss of generality, $\gamma$ is the self-intersecting curve given by
\begin{equation}\label{cos}
\cos\frac{n\vp}m=\chi(r^2,J^2,E)\,,
\end{equation}
where $\chi$ is the function
\[
\chi(r^2,J^2,E)=\begin{cases}
\displaystyle\frac{1+J^2\sqrt{r^{-2}+K}}{\sqrt{1+2J^2(E-G)+KJ^4}}\,\qquad &\text{if }\;j=\text{I}\,,\\[5mm]
\displaystyle
\frac{J^2r^{-2}\big(1-D r^2±\sqrt{(1-D r^2)^2-Kr^4}\big)+DJ^2+2G-2E}{\sqrt{(2E-2G-DJ^2)^2± 4J^2-KJ^4}}
 \, &\text{if }\;j=\text{II}\,.
\end{cases}
\]
Moreover, the chain rule immediately yields
\begin{align}
\sin\frac{n\vp}m&=-\frac m{n}\frac{\dd}{\dd\vp}\bigg(\cos\frac{n\vp}m\bigg)=-\frac {m\dot r}{n\dot\vp}\frac{\pd}{\pd r}\chi(r^2,J^2,E)=\Theta(r\dot r,r^2,J,E)\,,
\label{sin}
\end{align}
where
\[
\Theta(r\dot r,r^2,J,E)=-2r\dot r\frac {mr^2}{nJ}(D_1\chi)(r^2,J^2,E)
\]
and $D_1\chi$ stands for the derivative of the function $\chi$ with respect to its first argument. It should be noted that these expressions are well defined also for $J=0$.

Using the properties of the Chebyshev polynomials it is trivial to express $\cos n\vp$ and $\sin n\vp$ in terms of $r,\dot r,J$ and $E$ as
\begin{align*}
\cos n\vp&=T_m\bigg(\cos\frac{n\vp}m\bigg)=T_m\big(\chi(r^2,J^2,E)\big)\,,\\
\sin n\vp&=\sin\frac{n\vp}m\, U_{m-1}\bigg(\cos\frac{n\vp}m\bigg)=\Theta(r\dot r,r^2,J,E)\, U_{m-1}\big(\chi(r^2,J^2,E)\big)\,.
\end{align*}
Here $T_m$ and $U_m$ respectively stand for the Chebyshev polynomials of the first and second kind and degree $m$. Setting
\[
\SS^1=\big\{z\in\CC:|z|=1\big\}\,,
\]
we find it convenient to define the analytic $\SS^1$-valued map
\[
\cE_n(r\dot r,r^2,J,E)=T_m\big(\chi(r^2,J^2,E)\big)+\I \Theta(r\dot r,r^2,J,E)\, U_{m-1}\big(\chi(r^2,J^2,E)\big)\,,
\]
in terms of which the orbit $\gamma$ is characterized as
\begin{equation}\label{orbitexp}
\e^{\I n\vp}=\cE_n(r\dot r,r^2,J,E)\,.
\end{equation}

It stems from Fradkin's argument that~\eqref{Fradkin} yields a vector first integral of~\eqref{Hcoord} in any region where $\e^{\I\vp}$ can be unambiguously expressed in terms of the coordinates $(\bq,\bp)$. However, Eq.~\eqref{orbitexp} does not determine the angle $\vp$ univocally modulo $2\pi$ because the map $z\mapsto z^n$ of the unit circle onto itself has degree $n$, so that Fradkin's construction is, a priori, not global. As a matter of fact, it is obvious that Eq.~\eqref{orbitexp} only defines $\vp$ modulo $2\pi/n$, thus yielding an $n$-valued additional integral.

The aforementioned problem is a consequence of the fact that the orbit $\gamma$ has self-intersections. It is standard that this difficulty can be circumvented by means of an appropriate covering space of our initial manifold. The construction which we shall next outline is in fact analogous to that of the Riemann surface of the function $z\mapsto z^n$. We denote by $\gamma(t)$ the periodic integral curve of~\eqref{H} defined by the orbit $\gamma\subset M$ and take an $n$-fold cover $\Pi:\tM\to M$ of $M$ such that the lift $\tga(t)$ of $\gamma(t)$ to $\tM$ is a smooth path without self-intersections. Notice that $\tga(t)$ is actually an integral curve of the lifted Hamiltonian
\[
\tH_j=\frac12\big\|\widetilde p\big\|_{\Pi^*g_j}^2+(V_j\circ\Pi)(\widetilde q)\,,\qquad\J\,,
\]
where $(\widetilde q,\widetilde p)\in T^*\tM$. $\tM$ is a fiber bundle over $M$ with typical fiber $\ZZ_n$, and for each $k\in\ZZ_n$ we denote by $\La_k:M\to\tM$ the section of $\tM$ with fiber value $k$. Obviously $\La_k$ is an injective map, and an isometry from an open and dense subset $M_k\subset M$ onto its image in $(\tM,\Pi^*g_j)$. One obviously has that $\Pi\circ\La_k=\id$ and
\[
\Pi^{-1}(q)=\bigcup_{k\in\ZZ_n} \La_k(q)
\]
for all $q\in M$.

By construction, in each section $\La_k(M)$ there exists a determination of the (complex) $n$-th root which allows to solve $\e^{\I \vp}$ in terms of $\e^{\I n\vp}$ univocally along $\La_k(\tga)$. Therefore, for each $k\in\ZZ_n$ there exist real functions $S_k$ and $C_k$ (namely, determinations of $\arcsin$ and $\arccos$) such that
\[
\e^{\I \vp(t)}=C_k(\cos n\vp(t))+\I S_k(\sin n\vp(t))
\]
whenever the point $(r(t),\theta=\frac\pi2,\vp(t))$ lies in $\La_k(\tga)$. Moreover, an easy computation shows that the functions
\begin{align*}
\cC_k(r^2,J^2,E)&=C_k\big(T_m(\chi(r^2,J^2,E))\big)\,,\\
\cS_k(r\dot r,r^2,J^2,E)&=J^{-1}S_k\big(\Theta(r\dot r,r^2,J,E)\, U_{m-1}(\chi(r^2,J^2,E))\big)
\end{align*}
are analytic in their domains.

In order to express $\cC_k(r^2,J^2,E)$ and $\cS_k(r\dot r,r^2,J^2,E)$ in a more convenient way, we consider the lift of the coordinates $\bq$ to each space $\La_k(M)$. With a slight abuse of notation, we shall still denote these coordinates by $\bq$. An immediate computation shows that $\Pi^*g_j|_{\La_k(M)}$ reads as
\[
\dd s^2=\|\dd\bq\|^2+\big[h(\|\bq\|)^2-1\big]\frac{(\bq\cdot\dd\bq)^2}{\|\bq\|^2}\,,
\]
where the function $h$ is defined as in Section~\ref{S:potentials}, namely,
\[
h(r)^2=
\begin{cases}
\displaystyle\frac{m^2}{n^2( 1+Kr^2)}\qquad\quad&\text{if }\;j=\1\,,\\[5mm]
\displaystyle\frac{2m^2\left( 1-Dr^2±\sqrt{(1-Dr^2)^2-Kr^4}
\right)}{n^2\left((1-Dr^2)^2-Kr^4
\right)}&\text{if }\;j=\2\,.
\end{cases}
\]
By differentiation it stems from this formula that the conjugate momentum $\bp$ to $\bq$ is given by
\[
\bp=\dot\bq+\big[h(\|\bq\|)^2-1\big]\frac{\bq\cdot\dot\bq}{\|\bq\|^2}\,\bq\,,
\]
yielding $\dot\bq=\mathbf{v}(\bq,\bp)$ with
\begin{equation*}
\mathbf{v}(\bq,\bp)=\bp+\big[h(\|\bq\|)^{-2}-1\big]\frac{\bq\cdot\bp}{\|\bq\|^2}\,\bq\,.
\end{equation*}
As $r\dot r=\bq\cdot\dot\bq=\bq\cdot\mathbf{v}(\bq,\bp)$, we now have all the ingredients to invoke Fradkin's argument (cf.\ Eq.~\eqref{Fradkin}, with which~\eqref{Ak} should be compared) and derive that each component of the horizontal vector field $\bA_k$ in $T^*\La_k(M)$ defined by
\begin{equation}\label{Ak}
\bA_k=\frac1r\Big[\cC_k\big(\|\bq\|^2,\|\bq×\bp\|^2,H_j(\bq,\bp)\big)\,\bq+\cS_k\big(\bq\cdot\mathbf{v}(\bq,\bp),
\|\bq\|^2,\|\bq×\bp\|^2,H_j(\bq,\bp)\big)\,\bq×(\bq×\bp)\Big]
\end{equation}
is a constant of motion in $\La_k(M)$. By construction, the vector fields $\bA_k$ (with $k\in\ZZ_n$) define an analytic global horizontal vector field $A$ in $T^*\tM$ whose Lie derivative along the flow of $\tH_j$ is zero, thereby obtaining the desired unit Runge--Lenz vector.
\end{proof}

\begin{remark}\label{R:cover}
The particular form of the orbits~\eqref{orbits} and the fact that $\tM$ is a finite cover of $M$ ensure that all the lifted orbits which are bounded are also periodic, and that the lifted orbits do not have any self-intersections. Note that if $\tM$ is endowed with the pulled back metric $\widetilde g_j=\Pi^*g_j$, $\tH_j$ is a natural Hamiltonian system and $\Pi:(\tM,\widetilde g_j)\to(M,g_j)$ becomes a Riemannian cover.
\end{remark}

\begin{corollary}\label{C:n=1}
Consider a Hamiltonian of the form~\eqref{H} with $n=1$. Then the generalized Runge--Lenz vector is well defined in all $M$.
\end{corollary}
\begin{proof}
It trivially follows from Theorem~\ref{T:Runge}.
\end{proof}

\begin{corollary}\label{C:tensor}
Consider a Hamiltonian $H_j$ of the form~\eqref{H}, with $m,n$ coprime positive integers. Then there exists a horizontal symmetric tensor field in $M$ of rank $n$ which is invariant under the flow of $H_j$.
\end{corollary}
\begin{proof}
Let us use the same notation as in the proof of Theorem~\ref{T:Runge}. In particular, we consider the integral curve $\gamma(t)$ and the maps $\bA_k$ used in the proof of Theorem~\ref{T:Runge}. For each $k\in \ZZ_n$, let us denote by $\overline\bA_k(t)$ the restriction of the horizontal vector field $\bA_k:T^*\La_k(M)\to\RR^3$ to the projection of the integral curve $\gamma(t)$ to $T^*\La_k(M)$.
The only observation we need in order to prove Corollary~\ref{C:tensor} is that, by the expression for $\gamma$ found in Proposition~\ref{P:orbits} and the definitions of the covering space $\tM$ and of the horizontal vector fields $\bA_k$, it easily follows that
\[
\overline\bA_k\big(t+\tfrac \ell nT_\gamma\big)=\overline\bA_{k+\ell}(t)
\]
for all $k\in\ZZ_n$, $\ell\in\ZZ$, $t\in\RR$ such that $\gamma(t)\in M_{k+\ell}$ and $\gamma(t+\tfrac\ell nT_\gamma)\in M_k$. Here $T_\gamma$ stands for the period of the integral curve $\gamma(t)$ and the sum $k+\ell$ is to be considered modulo $n$. This periodicity property readily implies that the symmetric tensor product $\mathbf C$ of $\bA_1,\dots,\bA_n$, with components
\[
C^{i_1,\dots,i_n}(\bq,\bp)=A_1^{(i_1}(\bq,\bp)\cdots A_n^{i_n)}(\bq,\bp)\,,
\]
is a well defined, analytic tensor field in $M$ of rank $n$. As usual, symmetrization of the superscripts delimited by curved brackets is understood. To complete the proof of the corollary, it suffices to notice that $C$ is trivially invariant under the flow of $H_j$ as each $\bA_k$ is a (multivalued) first integral.
\end{proof}

Some comments may be in order. First, one should observe the dependance of the additional integrals~\eqref{Ak} on the momenta is generally complicated (and in particular not quadratic), which explains why they are usually so hard to spot~\cite{KKM06}. Second, it should be noticed that Corollaries~\ref{C:n=1} and~\ref{C:tensor} yield the usual Runge--Lenz vector and second rank conserved tensor (up to a normalization constant) when the Bertrand Hamiltonian we consider is the Kepler or harmonic oscillator system in Euclidean space~\cite{HM90,Pe92}. Note, however, that given an arbitrary Bertrand Hamiltonian it is usually hard to compute the conserved tensor $C$ or the Runge--Lenz vector $A$ in closed form. In this direction, it should be mentioned that an additional integral has been explicitly obtained for some of the Bertrand Hamiltonians discussed in Example~\ref{E: ex} (cf.\ e.g.~\cite{IK94,BHSS03,BEHR08a,Gi07} and references therein).

\section{Bertrand's theorem on curved spaces}
\label{S:examples}

In the previous sections we have thoroughly analyzed the superintegrability properties of the spherically symmetric natural Hamiltonian systems whose bounded orbits are all periodic. When combined with the discussion of harmonic oscillators and Kepler potentials on Bertrand spacetimes presented in Section~\ref{S:potentials}, this gives all the ingredients we need to state a fully satisfactory analogue of Bertrand's theorem on spherically symmetric spaces:

\begin{theorem}\label{T:main}
Let $H$ be the Hamiltonian function associated to a Bertrand spacetime, i.e., an autonomous, spherically symmetric natural Hamiltonian system on a Riemannian 3-manifold $(M,g)$ satisfying Properties (i) and (ii) in Bertrand's Theorem~\ref{T:Bertrand}. Then the following statements hold:
\begin{enumerate}
\item $H$ is of the form~\eqref{H} for some coprime positive integers $n,m$.

\item The potential $V$ is the intrinsic Kepler or oscillator potential in $(M,g)$.

\item $H$ is superintegrable. More precisely,
\begin{enumerate}
\item $H$ is geometrically superintegrable in the region of $T^*M$ foliated by bounded orbits.

\item There exists an $n$-fold cover $\tM$ of $M$ such that the lift of $H$ to $\tM$ admits a generalized Runge--Lenz vector.

\item There exists a nontrivial horizontal tensor field in $M$ of rank $n$ which is invariant under the flow of $H$.
\end{enumerate}
\end{enumerate}
\end{theorem}

As mentioned in the introduction, this result is of interest both in itself and because of the abundant literature devoted to the study of particular cases of this problem in different contexts and from various points of view.

\section*{Acknowledgements}

This work was partially supported by the Spanish Ministerio de Educación   under grant no.\    MTM2007-67389 (with EU-FEDER support)  (A.B.\ and F.J.H.), by the Spanish DGI and CAM--Complutense University under grants  no.~FIS2008-00209 and~CCG07-2779 (A.E.), and by the INFN--CICyT (O.R.).

\end{document}